\documentclass[a4paper,nobibnotes,nofootinbib,preprintnumbers]{revtex4}
\usepackage{amssymb}
\usepackage{amsmath}
\usepackage{epsfig}
\usepackage{subfigure,graphicx}

\newcommand{\be}{\begin{equation}}
\newcommand{\ee}{\end{equation}}
\newcommand{\bea}{\begin{eqnarray}}
\newcommand{\eea}{\end{eqnarray}}

\newcommand{\g}{\gamma}
\newcommand{\f}{\frac}

\newcommand{\br}{\mathbf{r}}

\newcommand\lr[1]{{\left({#1}\right)}}

\begin{document}

\title{Exclusive vector meson production at HERA from QCD with saturation}

\author{C. Marquet}\email{marquet@quark.phy.bnl.gov}
\affiliation{RIKEN BNL Research Center, Brookhaven National Laboratory, Upton,
NY 11973, USA}
\author{R. Peschanski}\email{pesch@spht.saclay.cea.fr}
\affiliation{Service de Physique Th{\'e}orique, CEA/Saclay,
91191 Gif-sur-Yvette cedex, France\\
URA 2306, unit{\'e} de recherche associ{\'e}e au CNRS}
\author{G. Soyez\footnote{on leave from the fundamental theoretical physics
group of the University of Li\`ege.}}\email{g.soyez@ulg.ac.be}
\affiliation{LPTHE, Universit\'e P. et M. Curie (Paris 6), Universit\'e D. 
Diderot (Paris 7)\\ Tour 24-25, 5e Etage, Bo\^ite 126, 4 place Jussieu, 75252 
Paris cedex 05, France\\
\hspace{1cm}UMR 7589, unit\'e mixte de recherche du CNRS\hspace{0.5cm}}

\preprint{SPhT-T07/026}
\preprint{RBRC-647}

\begin{abstract}
  Following recent predictions that the geometric scaling properties
  of deep inelastic scattering data in inclusive $\gamma^* p$
  collisions are expected also in exclusive diffractive processes, we
  investigate the diffractive production of vector mesons. Using
  analytic results in the framework of the BK equation at non-zero
  momentum transfer, we extend to the non-forward amplitude 
  a QCD-inspired forward saturation model
   including charm, following the theoretical predictions
  for the momentum transfer dependence of the saturation scale.
   We obtain a good fit
  to the available HERA data and make predictions for deeply virtual
  Compton scattering measurements.
\end{abstract}

\maketitle

\section{Introduction}

Geometric scaling \cite{gsincl} is a striking feature of deep
inelastic scattering (DIS) data at small Bjorken $x,$ or at large
rapidity $Y=\log(1/x).$ The photon-proton total cross section
$\sigma^{\g^*p\!\rightarrow\!X}_{\text{tot}}(x,Q^2)$ obeys a scaling
in the single variable $Q^2/Q^2_s(x)$ where $Q$ is the virtuality of
the photon and $Q_s$ the {\em saturation scale}. This momentum scale
increases with rapidity according to $Q_s^2(x)=Q_0^2\, x^{-\lambda}$
with $\lambda\!\sim\!0.3$ and $Q_0\!\sim\!0.1$ GeV (giving a
saturation scale of order 1 GeV for $x\!\sim\!10^{-4}$). Recently, the
diffractive cross-section
$\sigma^{\g^*p\!\rightarrow\!Xp}_{\text{diff}}$ and the elastic
vector-meson production cross-section
$\sigma^{\g^*p\!\rightarrow\!Vp}_{\text{VM}}$ were shown to also
feature scaling behaviours, with the same saturation scale
\cite{gsdiff}.

From the theoretical point of view, the high-energy or small$-x$ limit
can be studied within the QCD dipole picture \cite{dippic}.
Introducing the dipole-proton elastic scattering amplitude $T(r,Y),$
where $r$ is the transverse size of the dipole, the corresponding law
$T(r,Y)=T(rQ_s(Y))$ appears to be a genuine consequence of saturation
effects \cite{iim_gs} characteristic of the high-density regime of
perturbative QCD. More precisely, the evolution towards saturation is
conveniently described by the non-linear Balitsky-Kovchegov (BK)
equation~\cite{bk} that resums QCD fan diagrams in the
leading-logarithmic approximation. Geometric scaling for $T(r,Y)$
appears to be a mathematical consequence of the asymptotic solution in
$Y$ of this nonlinear equation in terms of travelling waves~\cite{tw}.
Within this theoretical framework, $Q_s(Y)=Q_0\Omega_s(Y)$ and the
rapidity dependence of the saturation scale, given by $\Omega_s(Y)$,
is then obtained from features of the
Balitsky-Fadin-Kuraev-Lipatov~\cite{bfkl} (BFKL) kernel, which drives
the linear regime of the BK evolution.

The total cross section $\sigma^{\g^*p\!\rightarrow\!X}_{\text{tot}}$
(as well as the diffractive cross section
$\sigma^{\g^*p\!\rightarrow\!Xp}_{\text{diff}}$ and the elastic
vector-meson production cross section
$\sigma^{\g^*p\!\rightarrow\!Vp}_{\text{VM}}$) is related to the
forward elastic dipole-proton amplitude. It is natural to ask the
question whether geometric scaling survives when considering
non-forward amplitudes. To that purpose, it is temptative to study the
BK equation which is written for $T(\textbf{r},\textbf{b};x),$ the
dipole-proton elastic scattering amplitude, depending not only on the
dipole size $\textbf{r}$ but also on the impact parameter
$\textbf{b}.$ However, the aforementioned scaling law cannot be easily
adapted to account for the impact parameter dependence. The analysis
of the BK equation in impact-parameter space~\cite{conf,bknum} even
shows a contradiction with confinement: the large-$\textbf{b}$
dependence of the solutions develops a power-law tail with decreasing
$x.$

Previous studies \cite{us,bkfull} have shown that the travelling-wave
method can be extended to the BK equation with transverse-space
kinematical dependence provided one investigates the problem in terms
of momentum transfer instead of impact parameter. The remarkable point
\cite{us} is that the travelling-wave analysis of the BK equation can
be better achieved in momentum space, by Fourier transforming
$T(\textbf{r},\textbf{b};Y)$ into $\tilde{T}(\textbf{r},\textbf{q};Y)$
where $\textbf{q}$ is the momentum transfer. From the knowledge of the
exact solutions of the BFKL equation at non-zero momentum
transfer~\cite{bfklsol}, the travelling-wave property has been
extended~\cite{us} (for large $Q$):
\begin{itemize}
\item at small momentum transfer, {\em i.e.} when
  $|\textbf{r}||\textbf{q}|<|\textbf{r}|Q_0<|\textbf{r}|Q\!\sim\!1,$ one has
  asymptotically
  $\tilde{T}(\textbf{r},\textbf{q};Y)\!=\!\tilde{T}(|\textbf{r}|Q_0\Omega_s(Y),
  \textbf{q})$, recovering the forward result;
\item at intermediate, semi-hard, momentum transfer, {\em i.e.} when
  $|\textbf{r}|Q_0<|\textbf{r}||\textbf{q}|<|\textbf{r}|Q\!\sim\!1,$ one has
  asymptotically
  $\tilde{T}(\textbf{r},\textbf{q};Y)\!=\!\tilde{T}(|\textbf{r}||\textbf{q}|
  \Omega_s(Y),\textbf{q});$
\item at large, hard, momentum transfer, {\em i.e.} when
  $|\textbf{r}|Q_0<|\textbf{r}|Q<|\textbf{r}||\textbf{q}|\!\sim\!1,$ one has
  no more saturation, (giving an explanation of the impact-parameter puzzle).
\end{itemize}
This introduces a $\textbf{q}$-dependent saturation momentum in the 
near-forward and intermediate transfer region with a
rapidity evolution $\Omega_s(Y)$ keeping the same form as in the
forward case.  Those predictions were confirmed both by analytical and
numerical analysis of the BK equation completely formulated in
momentum space~\cite{bkfull}.

Our purpose here is to investigate an interesting phenomenological
prediction from these results: the geometric scaling property should
manifest itself in exclusive vector meson production and deeply
virtual Compton scattering (DVCS), which are experimentally measured
 at moderate non-zero momentum transfer
$t=-\textbf{q}^2.$ In this paper, we analyse whether or not the
available data from HERA are sensitive to a $t-$dependent saturation
scale. To do this, we use a QCD-inspired saturation model for the
dipole amplitude $\tilde{T}(\textbf{r},\textbf{q};Y).$ Since it is
important to include the charm in the analysis, both for its impact
on DVCS, non-charmed mesons and also $J/\Psi$ production, we make use of a 
recent saturation model that successfully includes charm \cite{greg}. Note
that our parametrisation, which uses the momentum transfer $\textbf{q}$
instead of the impact parameter $\mathbf{b}$ as
suggested by this derivation of saturation effects in perturbative
QCD, provides a fruitful phenomenological framework since the data are
directly measured as a function of $t=-\mathbf{q}^2.$

The plan of the paper is as follows. In section II, we recall the
formulation of vector meson production and DVCS differential
cross-sections in terms of the dipole scattering amplitude
$\tilde{T}(\textbf{r},\textbf{q};x).$ In section III, we briefly
explain the asymptotic travelling-wave properties of this amplitude
and how they translate into geometric scaling at non zero momentum
transfer; we also introduce our QCD-inspired model for $\tilde{T}.$ In
Section IV, we present fits to the vector-meson production
experimental data, discuss our results and present predictions for
DVCS. Section V concludes.

\section{Exclusive vector meson production in DIS at small $x$}

\begin{figure}[t]
\begin{center}
\epsfig{file=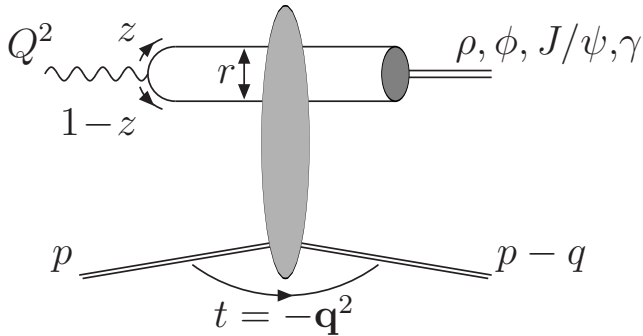,width=8.5cm}
\caption{Vector-particle production in the dipole frame. The amplitude
  factorises as a product of vertex functions and elastic
  dipole-proton interaction amplitude (see formula \eqref{eq:vmb_ampl}).}\label{fig:scheme}
\end{center}
\end{figure}

In the small$-x$ limit, it is convenient to describe the scattering of
the photon in a particular frame, called the {\em dipole frame}, in
which the virtual photon undergoes the hadronic interaction via a
fluctuation into a colorless $q\bar q$ pair, called dipole, which then
interacts with the target proton. The wavefunctions
$\psi^{\g^*,\lambda}_{f,h,\bar{h}}(z,\textbf{r};Q^2)$ describing the
splitting of a virtual photon with polarization $\lambda\!=\!0,\pm1$
into a dipole are well known. The indices $h\!=\!\pm1$ and
$\bar{h}\!=\!\pm1$ denote the helicities of the quark and the
antiquark composing the dipole of flavor $f.$ The wavefunctions depend
on the virtuality $Q^2,$ the fraction $z$ of longitudinal momentum
(with respect to the $\g^*\!-\!p$ collision axis) carried by the
quark, and the two-dimensional vector $\textbf{r}$ whose modulus is
the transverse size of the dipole. Explicit formul\ae\ for the QED
functions $\psi^{\g^*,\lambda}_{f,h,\bar{h}}$ can be found in the
literature~\cite{bjorken} and are recalled in Appendix \ref{app:wf}.

The exclusive production of vector mesons is represented in
Fig.\ref{fig:scheme}: the photon splits into a dipole of size
$\textbf{r}$ which scatters elastically off the proton, with momentum transfer
$\mathbf{q},$ and recombines into a vector meson whose
mass we shall denote $M_V.$ To describe this process, we need to
introduce the wavefunctions
$\psi^{V,\lambda}_{f,h,\bar{h}}(z,\textbf{r};M_V^2)$ which describe
the splitting of the vector meson with polarization $\lambda$ into the
dipole. In fact, to compute the vector-meson production amplitude
pictured in Fig.\ref{fig:scheme}, we need the transverse (T) and
longitudinal (L) {\it overlap functions}
$\Phi_T^{\g^*V}\!=\!(\Phi_+^{\g^*V}\!+\!\Phi_-^{\g^*V})/2$ and
$\Phi_L^{\g^*V}\!=\!\Phi_0^{\g^*V}$ obtained through 
\be\label{eq:wf}
\Phi_{\lambda}^{\g^*V}(z,\textbf{r};Q^2,M_V^2)
 = \sum_{fh\bar{h}} \left[
   \psi^{V,\lambda}_{f,h,\bar{h}}(z,\textbf{r};M_V^2)\right]^*
\psi^{\g^*,\lambda}_{f,h,\bar{h}}(z,\textbf{r};Q^2)\ .
\ee 
These functions depend on the meson wavefunctions
$\psi^{V,\lambda}_{f,h,\bar{h}},$ and different models exist
in the literature~\cite{mwfs1,mwfs2,mwfs3}. We shall discuss two
different choices later in this paper: the {\em boosted Gaussian} (BG)
wavefunctions \cite{mwfs1} and the {\em light-cone Gaussian}
(LCG) wavefunctions \cite{mwfs3}. For completeness, we give explicit
expressions for those overlap functions in Appendix \ref{app:wf}.

If $\textbf{q}$ denotes the transverse momentum transfered by the
proton during the collision, the differential cross-section with
respect to $t=-\textbf{q}^2$ reads 
\be 
\f{d\sigma^{\g^*p \rightarrow Vp}_{T,L}}{dt}
 = \f1{16\pi} \left[1+(\beta^{(V)}_{T,L})^2\right]\,
   \left| \mathcal{A}^{\g^*p \rightarrow Vp}_{T,L}\right|^2.
   \label{eq:vmb}
\ee
In that expression, $\mathcal{A}$ refers to the imaginary part of the
scattering amplitude and is given by
\be
\mathcal{A}^{\g^*p \rightarrow Vp}_{T,L}
 = \int d^2x\,d^2y\int_0^1 dz\
   \Phi_{T,L}^{\g^*V}(z,\textbf{x}\!-\!\textbf{y};Q^2,M_V^2)\
   e^{i\textbf{q}\cdot\textbf{y}}\ T(\textbf{x},\textbf{y};Y)\ ,\label{eq:vmb_ampl}
\ee 
where $\textbf{x}$ and $\textbf{y}$ are respectively the transverse
positions of the quark and the antiquark forming the dipole.
$T(\textbf{x},\textbf{y};Y)$ is the dipole-proton scattering amplitude
and carries all the energy dependence via the rapidity $Y$ which is
now obtained from the centre-of-mass energy $W$ and the vector-meson
mass $M_V$ using
\[
Y = \log\left(\frac{W^2+Q^2}{M_V^2+Q^2}\right)\ .
\]
The prefactor $1+\beta^2$ in \eqref{eq:vmb} accounts for the
contribution coming from the real part of the amplitude and can be
obtained using dispersion relations:
\begin{equation}
\beta_{T,L}^{(V)} = \tan\left(\frac{\pi\lambda}{2}\right)
\quad\text{ with }\quad
\lambda = \frac{\partial\log(\mathcal{A}^{\g^*p \rightarrow Vp}_{T,L})}
{\partial\,\log(1/x)}\ . 
\end{equation}

It is convenient to consider the dipole-proton scattering amplitude as
a function of $\textbf{r}\!=\!\textbf{x}\!-\!\textbf{y}$ and
$\textbf{b}\!=\!z\textbf{x}\!+\!(1\!-\!z)\textbf{y}$ and to introduce
the following Fourier transform:
\be
\tilde{T}(\textbf{r},\textbf{q};Y)=\int d^2b\
e^{i\textbf{q}\cdot\textbf{b}}\ T(\textbf{r},\textbf{b};Y)\ .  
\ee
Indeed, as we shall discuss in the next section, this quantity
features the geometric scaling property at non-zero momentum transfer.
The formula we shall use finally reads
\be 
\f{d\sigma^{\g^*p\rightarrow Vp}_{T,L}}{dt}=
  \f1{16\pi} \left[1+(\beta^{(V)}_{T,L})^2\right]\, \left|\int
  d^2r \int_0^1 dz\ \Phi_{T,L}^{\g^*V}(z,\textbf{r};Q^2,M_V^2)\
  e^{-iz\textbf{q}\cdot\textbf{r}}\
  \tilde{T}(\textbf{r},\textbf{q};Y)\right|^2\ .\label{eq:vmq} 
\ee

Note that formula \eqref{eq:vmq} can also be used to compute the DVCS
cross-section $\sigma^{\g^*p\rightarrow \g p},$ provided one uses the
vertex for a real photon instead of a vector meson in \eqref{eq:wf}.
The overlap function between the incoming virtual photon and the
outgoing transversely-polarized real photon is now model-independent
and given by \be \Phi^{\g^*\g}_T(z,\textbf{r};Q^2)=\sum_{fh\bar{h}}
\left[\psi^{\g^*,T}_{f,h,\bar{h}}(z,\textbf{r};0)\right]^*
\psi^{\g^*,T}_{f,h,\bar{h}}(z,\textbf{r};Q^2)\ .\label{dvcsof} \ee

\section{Geometric scaling at non-zero momentum transfer}

We have expressed the exclusive vector-meson production cross-sections
\eqref{eq:vmq} in the high-energy limit in terms of the
Fourier-transformed dipole scattering amplitude off the proton
$\tilde{T}(\textbf{r},\textbf{q};Y).$ Its evolution towards large
values of $Y$ is computable from perturbative QCD and the most
important result about the growth of the dipole amplitude towards the
saturation regime is probably the geometric scaling regime. It first
appeared in the context of the proton structure function, which
involves the dipole scattering amplitude at zero momentum transfer. At
small values of $x,$ instead of being a function of {\em a priori} the
two variables $r=|\textbf{r}|$ and $Y,$ the dipole scattering
amplitude is actually a function of the single variable $r^2Q^2_s(Y)$
up to inverse dipole sizes significantly larger than the saturation
scale $Q_s(Y).$ More precisely, one can write \be\label{eq:geomsc0}
\tilde{T}(\textbf{r},\textbf{q}=0;Y)=2\pi R_p^2\: N(r^2Q_s^2(Y))\ ,
\ee implying (for massless quarks) the geometric scaling of the total
cross-section at small $x:$ $\sigma^{\g^*p\rightarrow
  X}_{\text{tot}}(Y,Q^2)= \sigma^{\g^*p\rightarrow
  X}_{\text{tot}}(\tau\!=\!Q^2/Q_s^2(Y)).$ This has been confirmed by
experimental data~\cite{gsincl} (see also~\cite{gsdiff} for the
geometric scaling of the diffractive and elastic vector-meson
production cross-sections) with $Q_s^2(Y)=Q_0^2e^{\lambda Y},$ and the
parameters $\lambda\!\sim\!0.3$ and $Q_0\!\sim\!0.1$ GeV.

As explained in the Introduction, it has been shown in a recent work
\cite{us,bkfull} that the geometric scaling property can be extended
to the case of non zero momentum transfer, provided $r|\textbf{q}|\ll
1$. We obtained that equation \eqref{eq:geomsc0} can be generalised to
\be
\tilde{T}(\textbf{r},\textbf{q};Y)= 2\pi R_p^2 \:f(\textbf{q})\:
N(\textbf{r}^2Q_s^2(Y,\textbf{q}))\ ,
\ee
with the asymptotic behaviours $Q_s^2(Y,\textbf{q})\sim
\max(Q_0^2,\textbf{q}^2)\,\exp(\lambda Y)$ and an unknown form factor
$f(\textbf{q})$ of non-perturbative origin.

In practice, we need to specify three ingredients:
\begin{itemize}
\item The $\mathbf{q}$ dependence of the saturation momentum is
  parametrised as \be\label{eq:qst} Q_s^2(Y,\textbf{q})=Q_0^2(1+c
  \textbf{q}^2)\:e^{\lambda Y} \ee in order to interpolate smoothly
  between the small and intermediate transfer regions.
\item The form factor $f(\textbf{q})$ catches the transfer dependence
  of the proton vertex. It has to be noticed that this form factor is
  factorised from the projectile vertices and thus does not spoil the
  geometric scaling properties.  For simplicity, we use
  $f(\textbf{q})=\exp(-B\mathbf{q}^2)$.
\item The scaling function $N$ is obtained from the forward saturation
  model \cite{iim, greg}: 
\be \label{eq:t0}
N (rQ_s(Y),Y)=\left\{\begin{array}{lll} N_0\
    \lr{\f{r^2Q_s^2(Y)}4}^{\g_c}\
    \exp\left[-\f{\ln^2(r^2Q_s^2(Y)/4)}{2\kappa\lambda Y}\right]
    &\mbox{for }r^2Q_s^2(Y)\leq 4\\
    \\1-e^{-\alpha\ln^2(\beta\textbf{r}^2Q_s^2(Y))} &\mbox{for
    }r^2Q_s^2(Y)>4\end{array}\right.  
\ee 
with $\alpha$ and $\beta$ uniquely determined from the conditions
that $N $ and its derivative are continuous at $r^2Q_s^2(x)\!=\!4.$
The amplitude at the matching point is chosen to be $N_0=0.7.$

In this work, we shall consider the same form for the forward amplitude
as in the Iancu-Itakura-Munier (IIM) saturation model \cite{iim} but 
in its version extended in \cite{greg} to include heavy quarks (with $m_c\!=\!1.4\
\mbox{GeV},$ $m_b\!=\!4.5\ \mbox{GeV},$ and $m_f\!=\!0.14\ \mbox{GeV}$
for the light flavors). The coefficient $\kappa\!=\!9.9$ is obtained
from the BFKL kernel while the critical exponent $\g_c\!=\!0.7376$ is
fitted to the HERA measurements of the proton structure function,
along with the remaining parameters. The saturation scale parameters
are $\lambda\!=\!0.2197$ and $Q_0\!=\!0.298\ \mbox{GeV}$ and the
proton radius is $R_p\!=\!3.34\ \mbox{GeV}^{-1}.$ Note that, after
including heavy quark contributions, the saturation scale stays above
1 GeV for $x\!=\!10^{-5},$ rather than dropping to about 500 MeV as is
the case in other studies (see the discussion in \cite{greg}).
This parametrization is also successful in describing inclusive diffraction data \cite{cyr}.

The last factor in the expression for small dipole sizes introduces
geometric scaling violations, important when $Y$ is not large enough,
as predicted by the high-energy QCD evolution. It controls the way how
geometric scaling is approached. Indeed, it can be neglected for
$\log(r^2Q_s^2/4)<\sqrt{2\kappa\lambda Y}$, meaning that the geometric
scaling window extends like $\sqrt{Y}$ above the saturation scale (in
logarithmic units).
\end{itemize}

The final expression of our QCD-based saturation model is thus
\be\label{eq:T}
\tilde{T}(\textbf{r},\textbf{q};x)= 2\pi R_p^2\:e^{-B\textbf{q}^2}
N (\textbf{r}Q_s(x,\textbf{q}),Y)
\ee
which is an extension of the forward model \cite{iim,greg} including
the QCD predictions for non zero momentum transfer.
Indeed, this model reproduces the initial model for $\textrm{q}=0$ and
ensures that the saturation scale has the correct asymptotic
behaviours. We have two parameters: $c$ related to the scale at which
the $\textrm{q}-$dependence of the saturation scale becomes important
and $B,$ the $t-$slope of the form factor that we have taken as simple
as possible. Those parameters have to be fitted to the experimental
measurements of elastic vector-meson production as we shall comment in
details in the next section.

Note that the form factor depends on a single slope $B,$ independently
of the vector-meson in the final state. This is in agreement with
predictions from the BK equation \cite{us,bkfull} which implies the factorisation of the
non-perturbative contribution in \eqref{eq:T}. The difference in the
processes of $\rho,$ $\phi$ and $J/\Psi$ production (or even DVCS
cross-sections) fully comes from the vector-meson wavefunctions.
Through their wavefunction, different vector mesons are sensitive to
different dipole sizes, and thus feel saturation differently.

\section{Fit results and discussion}

In this section, we compare our formulation \eqref{eq:vmq} and
\eqref{eq:T} for vector-meson production with the experimental
measurements of $\rho,$ $\phi$ and $J/\Psi$ elastic production. We
shall first state explicitly the data used to perform the fit, then
give more details concerning the fit itself, present our results and
compare them with other possible approaches, and finally we shall give
predictions for DVCS measurements.

\subsection{Data selection}\label{sec:data}

In this analysis, we shall include the differential cross-section
$d\sigma/dt$ as well as the elastic cross-section
$\sigma_{\text{el}}.$ The dataset is obtained from the $\rho$-meson
production measured by H1 \cite{H1rho}, the $\phi$-meson production
measured by ZEUS \cite{ZEUSphi}, and the $J/\Psi$-meson production
measured by ZEUS \cite{ZEUSjpsi} and H1 \cite{H1jpsi}. We have not
taken into account the ZEUS data for $\rho$ mesons \cite{ZEUSrho}
since they lie in the low-$Q^2$ region where we do not expect the
condition $r|q| \sim\sqrt{|t|/Q^2} \ll 1$ to be valid. The ratio
$R=\sigma_L/\sigma_T$ between the longitudinal and transverse elastic
cross-sections is mostly dependent of the choice of wavefunction,
hence we have not included it in the analysis. Finally, this gives a
total of 269 data points which have to be reproduced with the two
parameters $B$ and $c$ in equation \eqref{eq:T}. Indeed, we keep the
same parameters as in \cite{greg} for $t=0$ as they provide a good
description of the inclusive $F_2$ data.

\subsection{Meson wavefunctions and alternative models}\label{sec:alternate}

In order to test the quality of our parametrisation with $t-$dependent
saturation scale, we shall compare our results with two alternative
parametrisations. Since the main aim of this work is to analyse the
$t-$dependence of the saturation scale, the first alternative we shall
consider is a model in which we only adjust the slope $B$ while
keeping the saturation scale independent of the momentum transfer.

For the second alternative, we notice that if one assumes a slope $B$
for the differential cross-sections at a given $W$ and $Q^2$, the
slope is experimentally observed to depend on $Q^2.$ Within our
parametrisation, it is the $t$ dependence of the saturation scale
which is expected to account for this behaviour in $Q^2.$ This will be
compared with a model where we keep $Q_s$ constant in $t$ ($c=0$) and
explicitly introduce a $Q^2$ dependence of the slope $B$. We shall
use $B(Q^2)=B+B'/(Q^2+M_V^2),$ which shows the same trend as the
experimental estimations of $B(Q^2).$ Note however that this last
two-parameter model is not compatible with the factorisation formula
\eqref{eq:vmq}, valid at small values of $x,$ where only the
wavefunctions depend explicitly on $Q^2$.

Moreover, in order to understand to what extend our analysis depends upon the
model used for the vector-meson wavefunctions, we shall test the
dependence of our results with respect to two choices which have
already proven to give good results. As mentioned before, we will
consider the boosted Gaussian (BG) and the light-cone Gaussian (LCG)
wavefunctions. For completion of the phenomenological discussion,
 we will also introduce some other combinations of wave-functions
called BLL and BLB, as explained in the next subsection.

\begin{table}[ht]
\begin{center}
\begin{tabular}{|c|c|c||c|c|c|c||c|c|}
\hline 
     &            &  & 
\multicolumn{4}{|c||}{$\chi^2/N_{pts}$ for $t$-dependent $Q_s$} 
&\multicolumn{2}{c|}{$t$-independent $Q_s$ for BLB case} \\
\cline{4-9}
VM  &   Observable     &  $N_{pts}$   & 
$\phantom{xx}$BG$\phantom{xx}$ & $\phantom{xx}$LCG$\phantom{xx}$ & $\phantom{xx}$BLL$\phantom{xx}$ & $\phantom{xx}$BLB$\phantom{xx}$ & 
$\phantom{xx}$$B$ = constant$\phantom{xx}$ & $\phantom{xx}$$B(Q^2+M_V^2)$$\phantom{xx}$ \\
\hline
\hline
$\rho$ & $\sigma_{\text{el}}$   & 47 & 0.984 & 1.486 & 0.888 & 0.936 & 1.182 & 1.745 \\
\cline{2-9}
       & $\frac{d\sigma}{dt}$ & 50 & 1.517 & 1.861 & 1.654 & 1.576 & 2.375 & 1.572 \\
\hline
$\phi$ & $\sigma_{\text{el}}$   & 34 & 1.999 & 1.304 & 1.323 & 1.327 & 2.732 & 1.175 \\
\cline{2-9}
       & $\frac{d\sigma}{dt}$ & 70 & 1.452 & 1.115 & 1.049 & 0.978 & 1.321 & 0.746 \\
\hline
$J/\Psi$ & $\sigma_{\text{el}}$ & 44 & 0.747 & 0.492 & 0.587 & 0.655 & 2.398 & 0.936\\
\cline{2-9}
       & $\frac{d\sigma}{dt}$ & 24 & 2.315 & 2.336 & 2.286 & 2.239 & 2.534 & 2.266 \\
\hline
\hline
\multicolumn{2}{|r|}{Total}   & 269 & 1.412 & 1.349 & 1.203 & {\bf 1.186} & 1.955 & 1.295 \\
\hline
\end{tabular}
\end{center}
\caption{$\chi^2$ results of the fits to the vector-meson production data. The different wavefunctions indicated (BG, LCG, BLL, BLB) are discussed in the text. For the best wavefunction (BLB), the $\chi^2$ values are compared with alternative models, that feature a $t-$independent saturation scale.}
\label{tab:chi2}
\end{table}

\begin{table}[ht]
\begin{center}
\begin{tabular}{|l|c|c|c|c|c|c|}
\hline
                 &\multicolumn{4}{|c|}{$t$-dependent $Q_s$}&         
\multicolumn{2}{|c|}{$t$-independent $Q_s$ (BLB)}       \\
\cline{2-7}
Parameter        &  BG  &  LCG  &  BLL & BLB  &  $B$=const.  & $B(Q^2+M_V^2)$  \\
\hline
$c$ (GeV$^{-2}$) & 4.077 $\pm$ 0.310
                & 4.472 $\pm$ 0.325
                & 4.258 $\pm$ 0.332
                & 4.041 $\pm$ 0.311
                &         -   
                &         -        \\
$B$ (GeV$^{-2}$) & 3.754 $\pm$ 0.095
                & 3.724 $\pm$ 0.093
                & 3.708 $\pm$ 0.097
                & 3.713 $\pm$ 0.096
                & 2.011 $\pm$ 0.031
                & 1.447 $\pm$ 0.043 \\
$B'$  &   -   &   -   &   -   &   -   &   -  
      & 4.245 $\pm$ 0.325 \\
\hline
\end{tabular}
\end{center}
\caption{Parameters obtained from the fits. 
The parameters of our model do not vary much
 when using different wavefunctions. The LCG
  wavefunction describes better the 
$\phi$ meson and the BG wavefunction describes 
better the $\rho$ meson. This is why the BLL or 
BLB choices give better global descriptions.}
\label{tab:params}
\end{table}

\subsection{Results}\label{sec:results}

\begin{figure}[t]
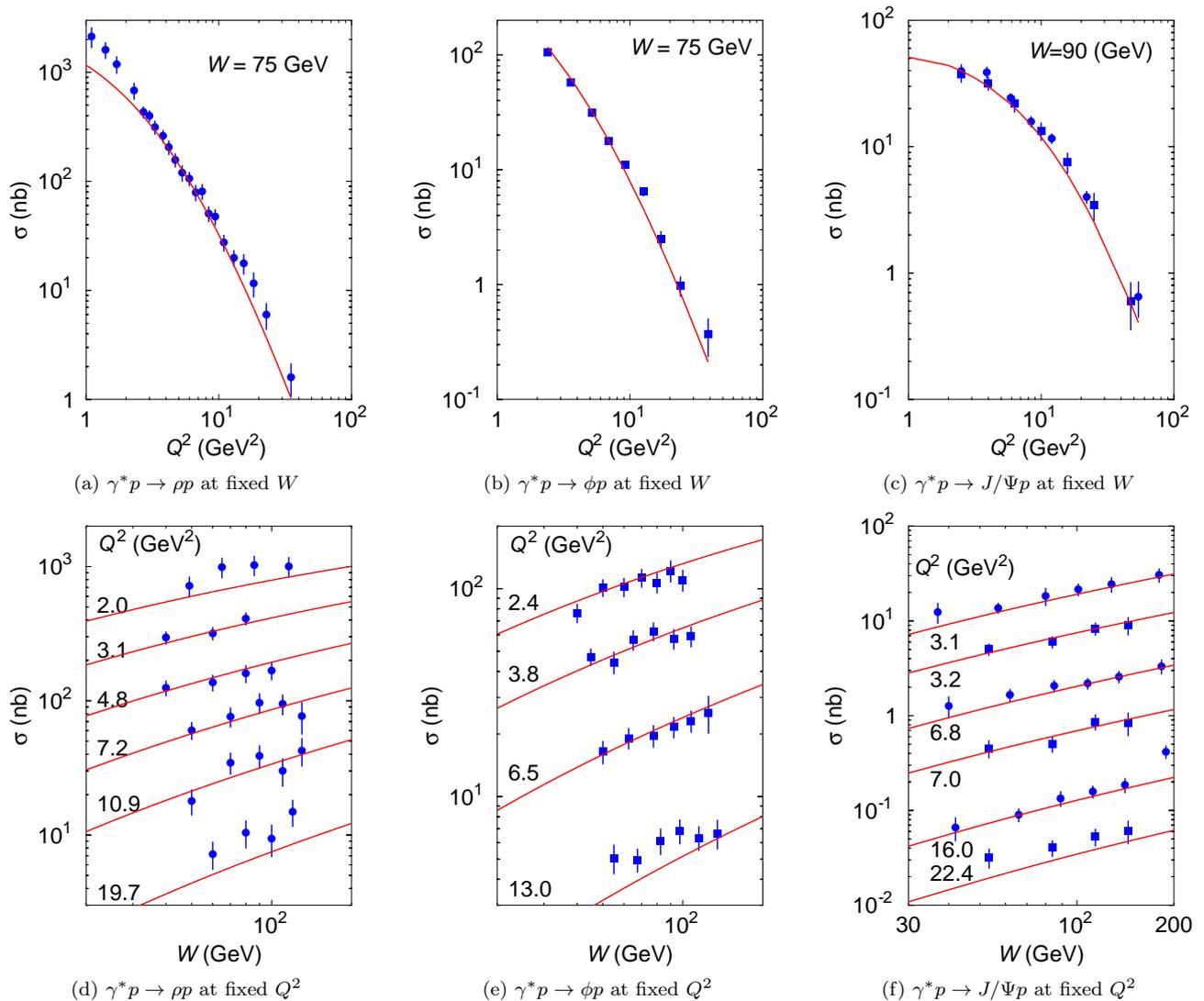

\begin{center}
 \subfigure[\ $\gamma^*p\to\rho p$ at fixed 
$W$]{\includegraphics[width=0.3\textwidth]{sigma_rho0.ps}}
\hspace{0.3cm}
 \subfigure[\ $\gamma^*p\to\phi p$ at fixed 
$W$]{\includegraphics[width=0.3\textwidth]{sigma_phi0.ps}}
\hspace{0.3cm}
 \subfigure[\ $\gamma^*p\to J/\Psi p$ at fixed 
$W$]{\includegraphics[width=0.3\textwidth]{sigma_jpsi2.ps}}

 \subfigure[\ $\gamma^*p\to\rho p$ at fixed 
$Q^2$]{\includegraphics[width=0.3\textwidth]{sigma_rho1.ps}}
\hspace{0.3cm}
 \subfigure[\ $\gamma^*p\to\phi p$ at fixed 
$Q^2$]{\includegraphics[width=0.3\textwidth]{sigma_phi1.ps}}
\hspace{0.3cm}
 \subfigure[\ $\gamma^*p\to J/\Psi p$ at fixed 
$Q^2$]{\includegraphics[width=0.3\textwidth]{sigma_jpsi1.ps}}
\end{center}
\caption{Fit results for the $\rho$ (H1 \cite{H1rho}), $\phi$ (ZEUS
  \cite{ZEUSphi}) and $J/\Psi$ (ZEUS \cite{ZEUSjpsi}) elastic
  cross-sections.}\label{fig:elas}
\end{figure}

\begin{figure}[ht]
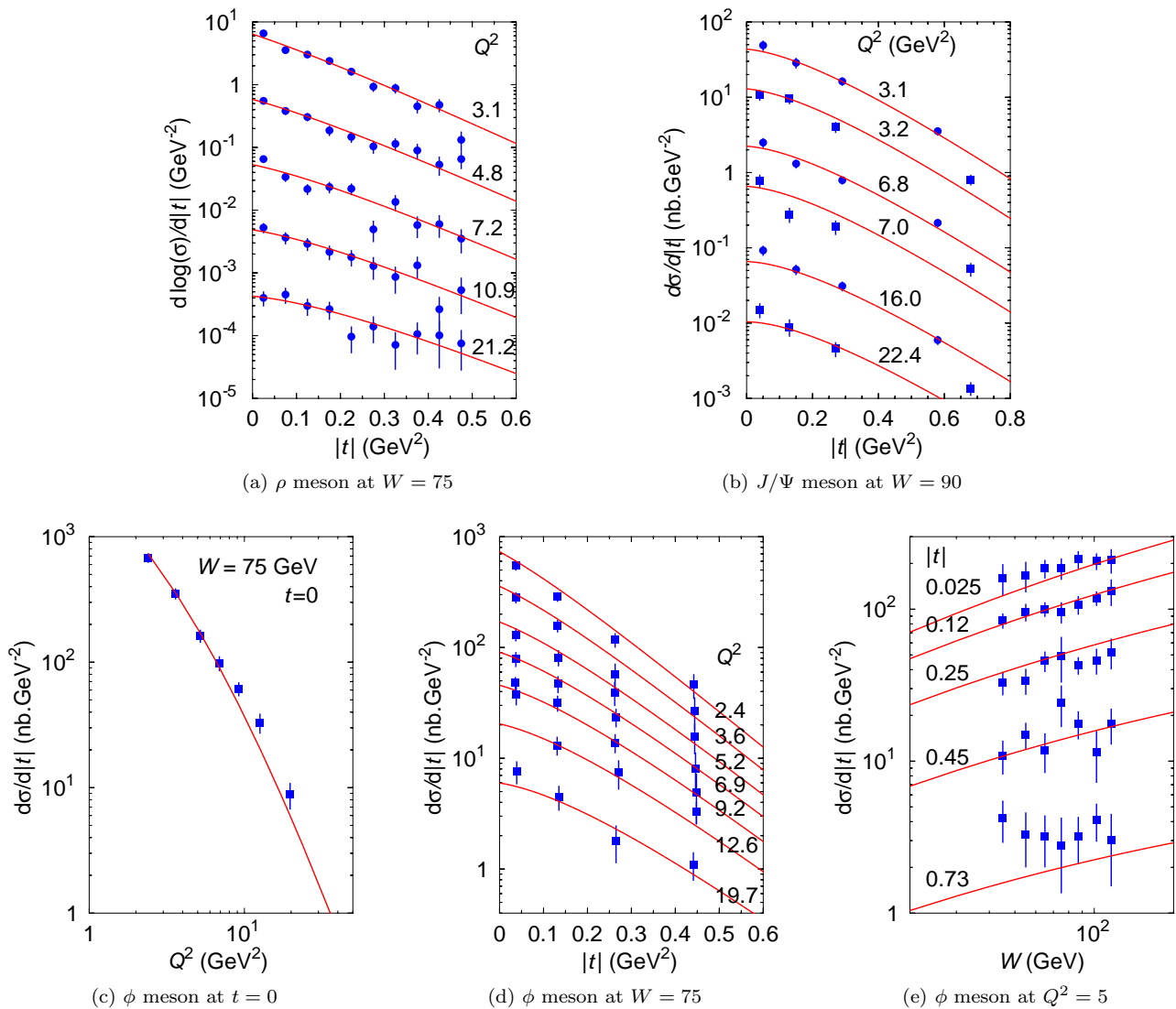

\begin{center}
\subfigure[\ $\rho$ meson at $W=75$]{\includegraphics[width=0.3\textwidth]{dsdt_rho.ps}}
\hspace{1.5cm}
\subfigure[\ $J/\Psi$ meson at $W=90$]{\includegraphics[width=0.3\textwidth]{dsdt_jpsi.ps}}

\subfigure[\ $\phi$ meson at $t=0$]{\includegraphics[width=0.3\textwidth]{dsdt_phi0.ps}}
\hspace{0.3cm}
\subfigure[\ $\phi$ meson at $W=75$]{\includegraphics[width=0.3\textwidth]{dsdt_phi.ps}}
\hspace{0.3cm}
\subfigure[\ $\phi$ meson at $Q^2=5$]{\includegraphics[width=0.3\textwidth]{dsdt_phi2.ps}}
\end{center}
\caption{Fit results for the $\rho$ (H1 \cite{H1rho}), $\phi$ (ZEUS
  \cite{ZEUSphi}) and $J/\Psi$ (ZEUS \cite{ZEUSjpsi}, H1
  \cite{H1jpsi}) differential cross-section.}\label{fig:ds_dt}
\end{figure}

We show in Table \ref{tab:chi2} the $\chi^2$ corresponding to the fits
of the different models presented above to the specified data. The
parameters associated with those fits are given in Table
\ref{tab:params}.  In both tables, successive columns show the results
for our model with $t-$dependent saturation scale convoluted with BG,
LCG, BLL and BLB wavefunctions (we shall discuss the wavefunctions
labelled BLL and BLB in a moment), followed by the two alternative
models without momentum-transfer-dependence in the saturation scale
(using BLB wavefunction). Those results show that our approach gives
good results for vector-meson production and this is confirmed by
Figures \ref{fig:elas} and \ref{fig:ds_dt} on which we have plotted
the curves for our fit together with the experimental data points.

\begin{figure}[ht]
\centerline{\includegraphics[angle=270,scale=0.7]{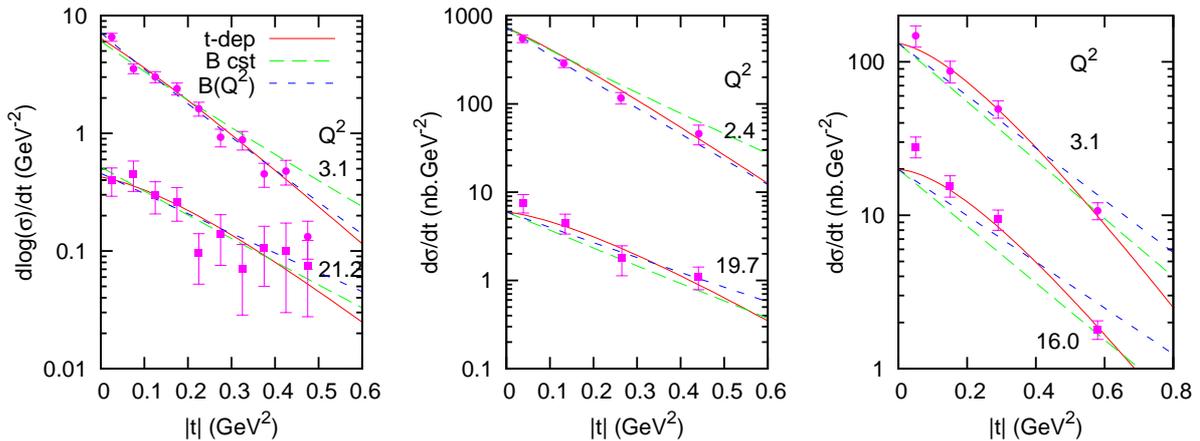}}
\caption{Comparison of the three parametrisations for differential
  cross-sections. The left plot shows the $\rho$-meson production at
  $Q^2$=3.1 and 21.2 GeV$^2,$ the middle one displays the $\phi$ meson
  at $Q^2$=2.4 and 19.7 GeV$^2,$ and the rightmost corresponds to the
  $J/\Psi$ meson at $Q^2$=3.1 and 16.0 GeV$^2.$ Continuous lines:
  $t$-dependent saturation; fat-dashed lines: $t$-independent
  saturation, fixed slope; thin-dashed lines: $Q^2$-dependent
  slope.}\label{fig:compare}
\end{figure}

\begin{figure}[ht]
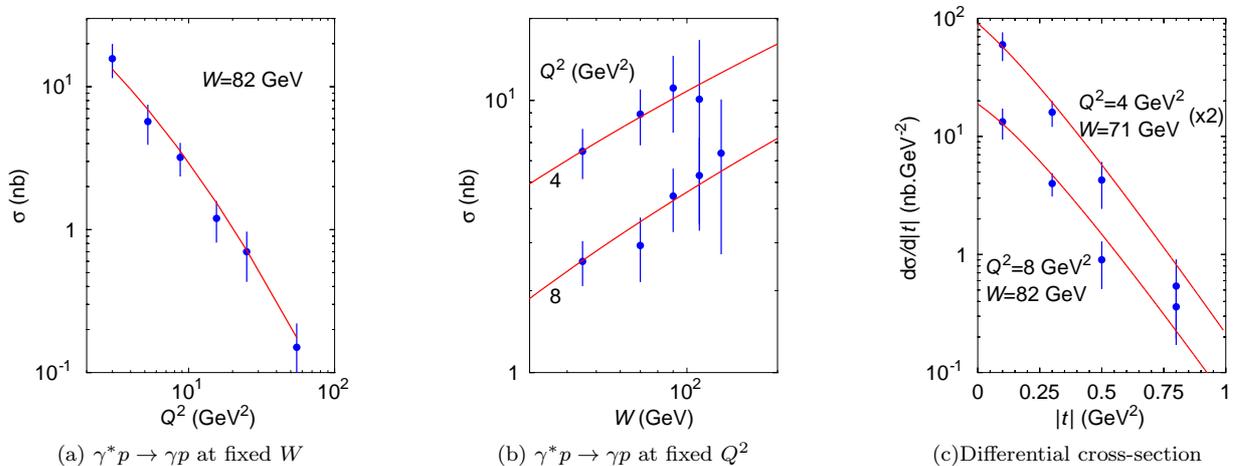

\begin{center}
  \subfigure[\ $\gamma^*p\to\g p$ at fixed
  $W$]{\includegraphics[scale=0.65]{sigma_dvcs0.ps}} \hspace{1cm}
  \subfigure[\ $\gamma^*p\to\g p$ at fixed
  $Q^2$]{\includegraphics[scale=0.65]{sigma_dvcs.ps}} \hspace{1cm}
  \subfigure[Differential
  cross-section]{\includegraphics[scale=0.65]{dsdt_dvcs.ps}}
\caption{Predictions for the DVCS measurements. The data are from H1
  \cite{H1dvcs}.}
\label{fig:dvcs}
\end{center}
\end{figure}

It can be realized from Figs.\ref{fig:elas} and \ref{fig:ds_dt} that
we obtain a satisfactory description of all data, considering the very
small number of free parameters.  It is of particular relevance for
the inclusion of charm in the analysis to obverve that the different
behaviour of the $J/\Psi$ cross-section at fixed $W$ from lighter
vector mesons is reproduced within the same model. Hence the predicted
mass dependence (via the rapidity dependence) together with the
transfer dependence of the saturation scale seem to be both relevant
in the successful description of data. Indeed, in our parametrisation,
the mass dependence is not present in the non perturbative form
factor.

Some technical comments are in order:
\begin{itemize}
\item First we notice that the $J/\Psi$ differential cross-section
  always features a bad $\chi^2,$ and this is the case for all the
  wavefunctions and models studied. This is probably due to a
  normalization discrepancy between the H1 and ZEUS data, which are
  both included in the fit (in all the cases under
    consideration, we obtain a good description of the ZEUS data and a
    rather poor description of the H1 data)	. Fortunately, this
  concerns only a few points, and it does not alter the global
  $\chi^2$ too much.
\item We obtain decent fits with both the BG and LCG wavefunctions
  though, if we have a closer look at the partial values of $\chi^2$, we realise
  that there is a significant dependence on the choice for the
  wavefunction. The LCG description is significantly better
  for the $\phi$ meson while the BG description is better in the
  $\rho$ meson case.  Globally, the LCG $\chi^2$ is slightly better
  than the BG one.

  This inspired the following studies: using simultameously the LCG
  wavefunction for the $\phi$ meson and the BG wavefunction for the
  $\rho$ meson. When the LCG wavefunction is used for the $J/\Psi$ the
  analysis is labeled BLL, and when the BG one is used, it is called
  BLB. 

  One sees that these combinations really improve the $\chi^2$ values
  (this is the case for all the three models under study). In
  addition, the parameters $c$ and $B$ do not depend much on the
  choice of the wavefunction. Hence, this represents a good global
  description of the data and shows that the confrontation of our
  model with the experiments is successful.

\item Comparing fits performed with our parametrisation \eqref{eq:T}
  and the corresponding one with a $t-independent$ saturation scale
  and constant slope, one sees that the introduction of $Q_s(t)$
  significantly improves the fit. This significative result, which is
  consistent with theoretical expectations, opens the way for further
  experimental test of the saturation regime of QCD with exclusive
  processes.

  Compared to our prametrisation, the parametrisation with a
  $t\!-\!independent$ saturation scale and with $B(Q^2)=B+B'/(Q^2+M_V^2)$
  shows $\chi^2$ values lower in the $\phi$-meson case but higher in
  the $\rho$- and $J/\Psi$-meson case. The global $\chi^2$ is slightly
  higher than that of our QCD-inspired model, in which this effective
  behaviour in $Q^2$ is accounted for by $t$ dependence of the
  saturation scale.

  In all cases, we have used the BLB choice for the wavefunctions as
  it gives the best results. By doing this we hope to minimise the
  impact of the wavefunctions in our discussion and focus on the
  description of the $t$ dependence of the scattering amplitude.

  For better comparison with the other parametrisations, we have
  displayed in Figure \ref{fig:compare} the curves corresponding to
  the three parametrisations of Table \ref{tab:chi2} (our
  $t$-dependent parametrisation for the BLB wavefunctions together
  with the two $t$-independent ones) for $\rho$-, $\phi$- and
  $J/\Psi$-meson differential cross-sections for the lower and higher
  available values of $Q^2$.

\item The plot on figure \ref{fig:ds_dt}(c) deserves a special
  comment. The data are for the differential cross-section at $t=0,$ therefore 
  the corresponding theoretical curves only depend on the parametrisation
  of the forward scattering amplitude, and on the choice for the $\phi$-meson
  wavefuntion. We have observed a $\chi^2$ per point of 3.157 (resp.
  1.625) for the BG (resp. LCG) wavefunction, suggesting again that
  the LCG wavefunction might be more appropriate to study the $\phi$ meson
  production.

\item From the value of the parameter $c\!=\!4$ GeV$^{-2}$ obtained in
  our approach, we can say that the saturation scale starts to
  increase with momentum transfer when $t \gtrsim c^{-1}\!=\!0.25$
  GeV$^2.$ It is interesting that this scale lies within the range of
  accessible data meaning that we can get insight on the dynamics of
  saturation by looking at the present HERA data.
\end{itemize}

\subsection{Predictions for DVCS}\label{sec:dvcs}

\begin{figure}
\centerline{\includegraphics[width=0.35\textwidth,angle=270]{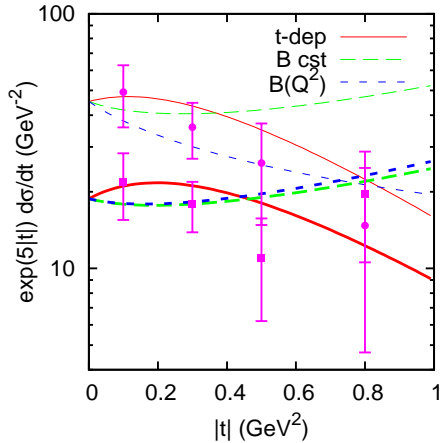}}
\caption{Predictions of the three models studied in this paper for the
  DVCS diffractive cross-section. For more clarity, we have multiplied
  the differential cross-section by $\exp(5|t|)$. The thin lines are
  related to $Q^2=4$ GeV$^2$ and $W=71$ GeV, while the thick ones are
  obtained with $Q^2=8$ GeV$^2$ and $W=82$ GeV.}
\label{fig:comparedvcs}
\end{figure}

Using the overlap function \eqref{dvcsof} in formula \eqref{eq:vmq},
we can predict the DVCS cross-section. Those predictions obtained with
our model and the parameters determined above are presented in Figure
\ref{fig:dvcs}. They are compared with a few available data
\cite{H1dvcs} and the agreement is good.  Because of the rather large
error bars, these data do not alter the fit described previously,
when included. Note that contrarily to diffractive vector-meson
production, the dipole-model description of the DVCS data involves a
wavefunction that can be computed from perturbative QED and does not
require a model. The DVCS predictions of the different models presented in the previous
sections differ quite a lot, at least in shape, as seen in Figure
\ref{fig:comparedvcs}. We hope that forthcoming DVCS data will
help to further test our predictions.

\section{Conclusions and perspectives}\label{sec:ccl}

Following the theoretical observation that the geometric scaling property
of the BK saturation equation is valid not only for inclusive but also
for diffractive exclusive processes of DIS, we have proposed an
extension of the parametrisation \cite{iim} (see formula \eqref{eq:T})
to non-forward vector-meson production.  The theoretical analysis
predicts a saturation scale proportional to $t$ and a factorisation of
the nonperturbative proton form factor. Hence, the geometric scaling
property, {\em i.e.} the $Q/Q_s(Y)$ scaling, is preserved at non-zero
momentum transfer with $Q_s(t,Y)=\Omega_s(Y)f(t)$, and $f(t)$
interpolating between the soft scale at small $t$ and $\sqrt{t}$ at
higher $t$.

By introducing two parameters to feature the predicted behaviour of
the saturation scale with $t$ (see \eqref{eq:qst}) and the factorised
non perturbative proton form factor (see \eqref{eq:T}), we satisfactory
describe the available HERA data for exclusive diffractive $\rho$-,
$\phi$- and $J/\Psi$-meson production.
This good agreement with the data shows the consistency between QCD
saturation predictions and measurements in the HERA energy range. We
shall be able to further test our parametrisation when the preliminary
$\rho,$ $\phi,$ and DVCS data become final. Note also that the
extention \eqref{eq:T} to non-forward amplitudes could be considered
for $t=0$ saturation parametrizations other than \eqref{eq:t0}.


Other successful models of exclusive processes with saturation effects
have already been achieved \cite{mms,kowtea,kmw,fss} confirming the
interest of saturation physics for vector-meson production.  However
we would like to emphasize the specific aspect of the present work,
which may open a new way to describe exclusive measurements in DIS.
Indeed, within our approach, we use the momentum transfer $\textbf{q}$
instead of the impact parameter $\mathbf{b}$ to parametrise the
saturation scale. This $\mathbf{q}$-dependence is expected
from p-QCD while the nonperturbative dependence in $\mathbf{q}$ is
factorised, which is not the case in impact-parameter space. This is
also practically convenient since the data are directly measured as a
function of $t=-\mathbf{q}^2.$


\begin{acknowledgments}

  We would like to thank Barbara Clerbaux, Laurent Favart, Alessia
  Bruni, Allen Caldwell and Miro Helbich for helping us collecting the
  HERA vector-meson production data. We also thank Laurent Schoeffel
  and Christophe Royon for insightful discussions concerning DVCS and
  Jean-Ren\'e Cudell for useful clarifications concerning the
  alternative models. C.M. is supported in part by RIKEN, Brookhaven
  National Laboratory and the U.S. Department of Energy
  [DE-AC02-98CH10886]. C.M. and G.S. also thank the Galileo Galilei
  Institute for Theoretical Physics for hospitality and the INFN for
  partial support when this work was completed. G.S. is funded by the
  National Funds for Scientific Research (Belgium). G.S. also wants to
  thank the SPhT (Saclay) for hospitality when this work was started.

\end{acknowledgments}
\eject
\appendix

\section{Photon and vector-meson wavefunctions}\label{app:wf}

For the self-consistency of the paper, let us give the expressions for
the overlap functions we have used in \eqref{eq:vmq} for the different
processes.  According to equation \eqref{eq:wf}, they are the product
of a wavefunction for the splitting $\gamma^*\to q\bar q$ and a factor
accounting for the production of the final state $q\bar q\to \gamma^*,
\gamma, V$, appropriately summed over the helicity and flavor indices.

When the final state is a photon (real or virtual), all vertices can
be computed from perturbative QED at lowest order in the
electromagnetic coupling. The results are
\begin{eqnarray*}
\Phi^{\gamma^*\gamma^*}_L(z,\br,Q^2) 
   & = &\sum_f e_f^2 \frac{\alpha_e N_c}{2\pi^2}\, 4 Q^2 z^2(1-z)^2 K_0^2(r\bar 
Q_f),\\
\Phi^{\gamma^*\gamma^*}_T(z,\br,Q^2) 
   & = &\sum_f e_f^2 \frac{\alpha_e N_c}{2\pi^2}\left\{
           [z^2+(1-z)^2]\bar Q_f^2 K_1^2(r\bar Q_f)
          + m_f^2K_0^2(r\bar Q_f) \right\},\\
\Phi^{\gamma^*\gamma}_T(z,\br,Q^2) 
   & = &\sum_f e_f^2 \frac{\alpha_e N_c}{2\pi^2}\left\{
           [z^2+(1-z)^2]\bar Q_f K_1(r\bar Q_f) m_f K_1(rm_f)
          + m_f^2 K_0(r\bar Q_f) K_0(rm_f) \right\},
\end{eqnarray*}
where $e_f$ and $m_f$ denote the charge and mass of the quark with flavor $f$ 
and with
$\bar Q_f^2 = z(1-z)Q^2+m_f^2.$ In practice, we sum over five flavors.

When the final state is a vector meson, the light-cone wavefunctions
are usually parametrised in terms of an additional unknown vertex
function for which there exists different models. The expressions to
be used in \eqref{eq:vmq} are then (for a vector-meson V)
\begin{eqnarray*}
\Phi^{\gamma^*V}_L(z,\br,Q^2) 
   & = & \hat{e}_f \sqrt{\frac{\alpha_e}{4\pi}} N_c \, 2QK_0(r\bar Q_f)  
\left[M_Vz(1-z)\phi_L(r,z)+\delta\frac{m_f^2-\nabla_r^2}{M_V}\phi_L(r,z)\right],\\
\Phi^{\gamma^*V}_T(z,\br,Q^2) 
   & = & \hat{e}_f \sqrt{\frac{\alpha_e}{4\pi}} N_c\frac{\alpha_e 
N_c}{2\pi^2}\left\{
             m_f^2 K_0(r\bar Q_f)\phi_T(r,z)
           - [z^2+(1-z)^2]\bar Q_f K_1(r\bar Q_f) \partial_r\phi_T(r,z)
            \right\},
\end{eqnarray*}
where the constant $\hat{e}_f$ stands for an effective charge. It is
given in table \ref{tab:wfparams} along with the quark and meson
masses used. Those expressions are very similar to the ones use for
photons except for the function $\phi_{L,T}$ that is related to the
vertex function and depends on the model. The first choice we have
investigated is the {\em boosted Gaussian} (BG) wavefunctions which is
a simplified version of the Nemchik, Nikolaev, Predazzi and Zakharov
model \cite{mwfs1} with $\delta=1$ and
\[
\phi_{L,T} = N_{L,T}
\,\exp\left[-\frac{m_f^2R^2}{8z(1-z)}+\frac{m_f^2R^2}{2}-\frac{2z(1-z)r^2}{R^2}
\right]\ .
\]
The parameters $R$ and $N_{L,T}$ are constrained by unitarity of the
wavefuntion as well as by the electronic decay widths. They are given
in table \ref{tab:wfparams}.

An alternative solution is the light-cone Gauss (LCG) wavefunction
\cite{mwfs3}. For those, we take $\delta=0$ and
\begin{eqnarray*}
\phi_L & = & N_L \,\exp\left[-r^2/(2R_L^2)\right],\\
\phi_T & = & N_T\,z(1-z)\,\exp\left[-r^2/(2R_T^2)\right],
\end{eqnarray*}
with the parameters also given in table \ref{tab:wfparams}.

\begin{table}[h]
\begin{center}
\begin{tabular}{|c||c|c|c||c|c|c||c|c|c|c|}
\hline
             & \multicolumn{3}{|c||}{common parameters}
             & \multicolumn{3}{|c||}{BG parameters}
             & \multicolumn{4}{|c|}{LCG parameters}\\
\cline{2-11}
Vector-meson & $M_V$ (GeV) & $m_f$ (GeV) & $\hat{e}_f$  
             & $R^2$ (GeV$^{-2}$) & $N_L$ & $N_T$ 
             & $R_L^2$ (GeV$^{-2}$) & $R_T^2$ (GeV$^{-2}$) & $N_L$ & $N_T$ \\
\hline
$\rho$  & 0.776 &  0.14 & 1/$\sqrt{2}$ 
        &  12.9 & 0.853 & 0.911 
        &  10.4 &  21.0 &  1.79 & 4.47 \\
$\phi$  & 1.019 &  0.14 & 1/3 
        &  11.2 & 0.825 & 0.919
        &   9.7 &  16.0 &  1.41 & 4.75 \\
$J/\Psi$& 3.097 &  1.4  & 2/3 
        &   2.3 & 0.575 & 0.578
        &   3.0 &   6.5 &  0.83 & 1.23 \\
\hline
\end{tabular}
\end{center}
\caption{Parameters for the vector-meson light-cone wavefunctions.}
\label{tab:wfparams}
\end{table}


\end{document}